# Are Some Technologies Beyond Regulatory Regimes?

Dimitri Kusnezov[1,2]

*US Department of Energy/NNSA, 1000 Independence Ave SW, Washington, DC 20585 USA*

Wendell B. Jones[3]

*Sandia National Laboratories, Albuquerque, NM, 87185 USA*

Regulatory frameworks are a common tool in governance to incent and coerce behaviors supporting national or strategic stability. This includes domestic regulations and international agreements. Though regulation is always a challenge, the domain of fast evolving threats, like cyber, are proving much more difficult to control. Many discussions are underway searching for approaches that can provide national security in these domains. We use game theoretic learning models to explore the question of strategic stability with respect to the democratization of certain technologies (such as cyber). We suggest that such many-player games could inherently be chaotic with no corresponding (Nash) equilibria. In the absence of such equilibria, traditional approaches, as measures to achieve levels of overall security, may not be suitable approaches to support strategic stability in these domains. Altogether new paradigms may be needed for these issues. At the very least, regulatory regimes that fail to address the basic nature of the technology domains should not be pursued as a default solution, regardless of success in other domains. In addition, the very chaotic nature of these domains may hold the promise of novel approaches to regulation.

Interesting questions have emerged in recent years asking how to understand regulation and strategic stability in domains where we see barriers to deploying novel technologies eroding globally (see, e.g., Lehman 2014). The discussion has also been framed as deterrence and dissuasion of adversaries using these technologies (Nye 2017). This "technology democratization" includes developments in manufacturing, computing/cyber/Internet of Things (IoT)/artificial intelligence (AI), chemistry, biology/gene editing/CRISPR-Cas9 and so forth, but also technologies that arise from the rapid convergence across these areas. For example: What do import/export control regulations mean in an era of 3D printing? What do international agreements on gene editing mean when kits to edit DNA can be ordered on Amazon, eBay, or Alibaba? Many of these technologies bring about deep questions on how to impose ethical standards, develop overarching regulatory structures, or whether to push for the development of treaties and other international agreements.

---

It is appealing to refer to historical approaches that have provided periods of strategic stability in other domains. Numerous efforts are underway to do just this. We will try to understand whether these experiences and successes are applicable to emerging areas driven by these technologies. Among these experiences are treaties tied to nuclear testing, such as the Threshold Test Ban Treaty (TTBT), those tied to non-proliferation of weapons, such as the treaty on the Non-Proliferation of Nuclear Weapons (NPT), or regimes that address the control of materials, tools and approaches such as the Wassenaar Arrangement, or the Nuclear Suppliers and Australia Groups. There has been the sense that properly phrased treaties or regimes could have similar desired effects for regulating these democratized technologies if we know how to better structure them. The view has been that these domain are quantitatively different, but not qualitatively different. This is the issue we will be exploring.

In a multilateral world, traditional approaches to strategic stability may not be successfully extended to encompass technologies of mass empowerment such as cyber or gene editing. For cyber, outside of national efforts, a few multinational approaches have been championed, such as the Budapest Convention. At the same time, the pace of the technologies, accelerating through the Internet of Things, IoT, and global interconnectivity for the case of cyber but analogous for other technologies, is proving challenging for such regulatory systems to accommodate. In the case gene editing, little beyond imputing ethical norms on practitioners is currently being done (NAS 2017; Evitt, Mascharak, and Altman 2015). We would like to understand more fundamentally whether the approaching stability using traditional approaches will have any impact on areas driven by these heavily democratized technologies.

To approach this topic, we would like to appeal to an analytic construct that allows for a more robust exploration of qualitative differences between traditional approaches to stability and the nature of democratized technologies. Game theory approaches can serve such a purpose. While having earlier roots, the bipolar dynamics of the Cold War helped blossom this field and its application (Poundstone 1992). As a tool, it provides a means to understand strategies and options and has found application to a broad range of fields (Camerer 2003). Within all these applications, an important concept is the Nash equilibrium (Nash 1950). This reflects a stable outcome where those involved in the defined game, the players, are aware of each other's strategies, and recognize that any deviations from where they currently are does not yield any benefit to themselves. The Nash equilibria in games are inherent to each specific game, and players over time can converge to those states. We could say that our shared experience in the negotiation of treaties and regulatory regimes historically could be put into the context of the game theoretical approach as follows.

The purpose of the negotiations can be viewed as the development of a shared awareness of the characteristics of the choice-space facing the parties in the negotiation. In a successful negotiation, the parties find an agreement that represents the optimal trade-off for every party. By definition, these are Nash equilibria. The existence of these Nash equilibria is due to the structure and context of the game, not to the negotiations themselves. The negotiators deserve credit for the exploration and discovery of the Nash equilibria, but they did not create them. Rather the (international) negotiations help to non-violently explore the space of available strategies and identify the potential Nash equilibria. The outcome of a sustainable agreement - and strategic stability - in this sense could be parsed as the mutual discovery of a Nash equilibrium in the game.

This perspective could also be used to establish if a domain is amenable to a negotiated regulatory framework. If an underlying system or game has Nash equilibria, then a regulatory regime would be the mechanism to create a shared awareness that illuminates the characteristics of the game. There are many places where such regimes have yielded long term international stability. In turn, successful regulatory regimes in the emerging domains can be seen as contingent on the existence of a Nash equilibrium. Then, and only then, can negotiations produce and enduring regulatory framework.

We will now examine these emerging domains from this perspective by reviewing some of the salient features of technologies that are increasingly democratized. The traditional role of territorial boundaries as organizing principles for governance are giving way to global interconnectedness and a concomitant technology empowerment . Tied to this are the evolving and emerging interdependencies of technologies as well as short time-scales for evolution compared to that of developing regulatory frameworks. The technology empowerment that is characteristic includes the increased accessibility and reduced costs. Open platforms enable open innovation and enable "fast followers" and co-innovators. Such empowered actors develop in turn other surprises, increasing pace and magnitude of potential disruptions. Society (and regulators) have difficulty keeping pace with rapid advances defined by rapid learning and a large and growing number of participants.

This type of fluid environment, with growing and evolving power centers, increases the complexity of the challenges associated with regulatory regimes and the uncertainties in the outcomes. It may not be possible to regulate heavily democratized technologies in the traditional sense. Even the international variations in how local or regional regulatory regimes are developed may render local standards ineffective, drive research to less restrictive countries, and create asymmetries in capabilities. The rapid democratization requires new approaches when technological superiority is not viable and responses become largely reactive and often too late, especially with the traditional R&D cycle.

In recent years, efforts to include learning and other characteristics into game theory have led to more realistic models of human behavior that we could use to capture some of the elements described above of those empowered to use democratized technologies (players). These include large numbers of players, with many diverse motives and strategies and learning that could be paced with technology and with correlations among the players. Behavioral models are reasonable approaches since they can begin to capture the dynamic learning that evolves with technologies and that player strategies can evolve in time based on such experiences and the communications that exist between them. Among these approaches, we will focus on the Experience Weighted Attraction (EWA) models of Camerer and Ho (Camerer and Ho 1999; Camerer 2004). Recent analyses of these models explored and characterized their dynamical behaviours (Galla and Farmer 2013; Sanders, Galla, and Farmer 2016), allowing the tools and methods from the field of complex physical systems to be applied. We will not attempt to reproduce the full details of EWA models since these are well described in the literature. Rather we will focus on some of the main features and some recent results. For the purpose of the qualitative analysis below, we will focus on cyber, where the players are those who can access relevant information technologies, who have suites of options on what they may do with these technologies given the backdrop of the growing IoT, and where the strategies dynamically evolve apace with technology. With the large body of work that exists in cybersecurity, including in both regulatory approaches and national and international responses, game theory in many forms has been used as a tool to help understand the response of players, however along much more traditional lines (Mikolic-Torreira et al. 2016).

This analysis utilizes the nomenclature developed for describing aspects of game theory. The core aspects for any game are: Who are the players? What are the choices (strategies) available to each? And, what payoffs (positive or negative) accrue from each choice by each player? For a complete description of this field, see *Behavioral Game Theory: Experiments in Strategic Interaction* by Colin Camerer (Camerer 2003).

In the EWA models, a number of players p each have strategies they can exercise, $x_i^\mu$, where $\mu = 1,\ldots, p$ identifies each player and $i = 1, \ldots, N$ reflects each of the *N* strategies available to each player. Strategies and players should be broadly interpreted as any application of technologies at the disposal of individuals, groups, or other entities including nation states. The strategies of the players evolve in time through learning and communication with other players. In the notation of (Sanders, Galla, and Farmer 2016), they are expressed as:

$$x_i^\mu(t+1) \propto \exp[\beta Q_i^\mu(t)] \qquad (1)$$

where β reflects how much current choices are effected by prior successes (Sato and Crutchfield 2003) and $Q_i^\mu(t)$ are dynamic and reflect the 'attractions' to a given player μ with strategy i. Numerically, a large value for β represents players that are driven by choosing the most rewarding strategy and simply repeat it over and over, while small β reflects players who tend to choose strategies randomly in time. The exponential form has been empirically justified (Camerer and Ho 1999), and also allows for invariance with respect to a constant shift in $Q_i^\mu(t)$ and for positive and negative attractions.

The attractions are updated in time according to (Sanders, Galla, and Farmer 2016):

$$Q_i^\mu(t+1) = (1-\alpha)Q_i^\mu(t) + \sum_{\{-i\}} \Pi_{i,\{-i\}}^\mu \prod_{\kappa \neq \mu} x_{i_\kappa}^\kappa(t) \qquad (2)$$

In this form there are two competing terms: a "memory loss rate" α and the correlation or "payoff" matrix $\Pi_{i,\{-i\}}^\mu$. The memory loss rate α, which ranges from 0 to 1, captures how much of prior experience and success in applying the technologies continues to impact decisions at later times. For smaller α, there is a stronger memory of prior history of successful uses of the technologies. For α closer to 1, there is smaller impact on prior applications of the technologies on what a given player will do next. The risks and rewards for a player's actions through the choice of their strategy is captured in payoff matrix $\Pi_{i,\{-i\}}^\mu$, which denotes the payoff to player μ if they play strategy *i* and the other players play all their respective strategies, denoted simply as $\{-i\}$. The symbol $\{-i\}$ is intended to be short-hand to represent the strategies of all other $(p-1)$ players other than player *i*, and their corresponding strategies. We do not attempt to recreate the full details of these models found in (Camerer 2004; Sanders, Galla, and Farmer 2016) and only wish to highlight the structures, their relationship to democratized technologies and potential dynamical ranges for the behaviors. Equation (2) balances how much a strategy is impacted from learning versus interactions and correlations with the other players who have their own strategies.

For a heavily democratized technology, an abundance of players with many strategies available to them, and in the absence of more specific information on the players, it is reasonable to apply a maximum entropy principle to the structure of the payoff matrix, as in (Berg and Weigt 1999; Sanders, Galla, and Farmer 2016). The overall statistical strength of interactions between players then has the following gross structure:

$$\langle \Pi^{\mu}_{i_{\mu},\{-i_{\mu}\}} \Pi^{\nu}_{i_{\nu},\{-i_{\nu}\}} \rangle \propto \Gamma + \delta_{\mu\nu}(p - 1 - \Gamma) \qquad (3)$$

The correlation between different players $(\mu \neq \nu)$ is characterized by a strength $\Gamma$, which can range from *-1* to the number of strategies *p-1*. In this notation, the impacts on the payoffs range from fully correlated players in a zero-sum game ($\Gamma = -1$), to the opposite limit in which players are fully uncorrelated with each other ($\Gamma = 0$), or to the situation where all payoffs are all identical regardless of what the players do ($\Gamma = p - 1$).

What is important here is the sign of the correlation function, $\Gamma$, and how players cooperate. In this elementary instantiation of an EWA model, we have five relevant parameters: $\alpha \in [0,1], \beta \in [0,\infty), \Gamma \in [-1, p-1]$, the number of players *p* and the number of strategies *N* per player.

There are many ways to think of the correlation function for the payoffs. With democratized technologies, by definition, there are many players and the empowerment is possible because of the correlations between players through awareness of technological opportunities and surprises. One could view the vitality of the Dark Web as an illustration of how players might wish to retain correlations among themselves but not necessarily cooperate in a specific shared goal. There are exceptions of course. In the case of the members of the group calling itself Anonymous, there was an appeal to define a shared outcome and payoff, but largely these are exceptions (Firer-Blaess 2016). So we might expect that $\Gamma \in (-1,0)$ would be a good characterization for the cyber domain: correlated, competitive, not fully cooperative and not zero-sum. For learning, we would expect that players learn from prior attempts to apply strategies, so that α would be closer to *0* than *1*. Similarly, if we expect players to use strategies that are not fully random but rather built on prior successes, then we expect β to be a large positive quantity. Hence it is likely that for cyber as well as democratized technologies, $0 \lesssim \alpha \ll \beta$, and both *N* and *p* are large positive integers.

In the study of complex physical systems, stability analysis helps characterize the expected nature of the system evolution over time. Stability analysis of particular limits of EWA models have recently found that there can be regions without Nash equilibria and that these are characterized by chaotic responses by the players. That is to say, players respond to the structure of the payoff (risk/reward) structure in unpredictable ways. Our experience in the study of complex systems, both at the classical and quantum mechanical level, is that the presence of chaos is commonplace, and its presence allows for extraction of generic features of theories and of nature (Reichl 2004). We would expect that generic features of chaotic many-person game theory models likely exist and could develop in time. In few-body physical systems, as in few-player games (i.e. the Cold War), the absence of chaos is typical. However, it does not take too many additional degrees of freedom before physical systems get chaotic, often due to non-linearities and/or feedback, and as a matter of course, chaotic behavior is more the norm than the exception (Skyrms 1992). In chaotic games, player responses become dynamically unstable and unpredictable, and as a consequence the game itself may not have any (Nash) equilibria. We would like to use the structure of these many-player games to better define differences between historic regulatory regimes and the challenges we face today with the democratization of certain technologies. This would help guide work hoping to produce stability in these important domains.

Within the analysis of EWA models, and in the limit of a large number of players each with many strategies, the boundary between stable (non-chaotic) and unstable (chaotic) regimes of player responses can be extracted from (Sanders, Galla, and Farmer 2016) to be approximately given by:

$$\frac{\alpha}{\beta} \sim \sqrt{p} \quad . \tag{4}$$

So for $\frac{\alpha}{\beta} \gtrsim \sqrt{p}$ the behaviour is not chaotic and player strategies converge to stable equilibria – that is, the systems evolves into a controlled outcome for behaviors. But for smaller values and $\frac{\alpha}{\beta} \lesssim \sqrt{p}$ the behaviour of the system is chaotic and the response of players is unstable and they will evolve in time effectively randomly despite the controls on their behaviour. As we have argued for cyber, $0 \lesssim \alpha \ll \beta$, so that $\alpha/\beta \ll 1$ is quite reasonable so that behaviour of the system is well within the chaotic domain and is without any equilibria. Consequently, if one wanted to devise a regulatory regime that would drive to stable outcomes in the more traditional sense, one would need to satisfy:

$$\beta \lesssim \alpha/\sqrt{p} \tag{5}$$

Keeping in mind that $\alpha \in [0,1]$, and the number of players is large by definition, one would need regulatory frameworks to drive players to $\beta \to 0$, or compelling players to weigh all of their available strategies with equal probability of use. One can view this as heavy coercion that makes any play equally risky or poor, regardless of their experience and prior successes, or it could be viewed as appealing to their good sides and adherence to ethical standards by a preponderance of players.

In the absence of such sweeping mechanisms to modify behaviors, the outcome will be the absence of Nash equilibria, and ongoing chaotic behaviors. But again, regardless of the overarching regulatory construct, there are no points of stability. This latter case could be viewed as the implementation of dissuasion through norms, as proposed by (Nye 2017).

The utility of Cold War Era game theory was facilitated by the very limited number of players (nuclear capable nations). With the extraordinarily high entry cost to become a nuclear power, the conditions for using game theory were tractable and that approach has proved useful. These conditions are not present when looking at such highly democratized technologies as cyber and gene editing. The more recent development of many-player learning models in game theory can provide a means to capture some of these salient features of the complex global ecosystem of players that have growing access to increasingly democratized technologies.

What does chaos mean? It is the time evolution of player strategies that are found to evolve chaotically in time or alternatively, player responses are unpredictable and there is no equilibrium to be had. In the study of physical systems, the identification of chaotic behavior has allowed use of techniques from dynamical systems theory or random matrix theory to compute quantities that could then be related to experimental observations. As a general rule, many physical models with realistic features are analytically intractable, so that chaotic or ergodic limits of their behavior can be used as a means to identify behaviors that are robust and independent of model specifics. It also allows the computation of certain types of physical observables that behave generically or universally. The identification of regimes where physical systems are chaotic has allowed the prediction and understanding of many experimental observations.

Game theory based models remain as abstractions of human behaviors, but with increasing refinements and validation over time they can be useful tools to help frame thinking and actions. Looking at this class of problems today in a more rigorous analytic approach may not shed further light. But it does lead to classes of questions we should be asking as we approach ubiquitous technologies that have dual use characteristics. While we focused on cybersecurity in this analysis, democratized technologies such as gene editing (CRISPR-Cas9 and its follow-ons) share the same characteristics. These are technologies where the technical barriers for entry and the associated costs and infrastructure continue to see dramatic reductions so that traditional international variations in approaches to such rapidly changing fields can render local standards ineffective and drive activities to less restrictive places. Within the fields themselves there can be growing recognition that action is needed. In the regulatory construct recently proposed for gene editing technologies (Evitt, Mascharak, and Altman 2015), we would argue that this approach provides ethical people a means to be ethical and demonstrate that they are. Approaches like this can bias good people to be good (e.g. through laws, penalties, etc) but they don't change the underlying state of the system. The recent study by the National Academies have also weighed in on ensuring adherence to ethical standards (NAS 2017). The question for any of the democratized technologies is how does one drive this to $\beta \rightarrow 0$ in order to satisfy Eq. (5)?

This analysis indicates that it is possible that chaotic games will lead to classes of model-independent predictions that help understand the features of the complex human behaviors that the games are meant to explain (Chakraborti et al. 2015). This understanding could, in turn, guide policy and regulatory negotiations in the most fruitful directions. It is possible that salient features are expressible through such models and, as a consequence, help in guiding new approaches to dealing with heavily democratized technologies could emerge. This is true even though we cannot expect game theory to be detailed predictors of human behavior. Rather it may be a first step to recognize that something different may be needed. Finally, it is worth noting that in chaos theory, the field of controlling chaos emerged as a means to provide (small and well placed) feedback into a chaotic system to limit instabilities in the behavior of a system (Ott, Grebogi, and Yorke 1990). If regulatory regimes for democratized technologies fall into the category of chaotic games, as we argue, it might be that there are analogous types of feedback that could be developed to mitigate the resulting unpredictability of behavior. Suitable feedback in broader definitions of regulatory regimes or even preparedness against threats coevolving with technology could be a parallel to bringing some form of strategic stability. Whether or not we choose EWA models as a starting point, the democratization of technologies by definition is underpinned by a breadth of players, many targets and strategies, and a complex risk/reward structure. If learning is part of the player experience, and players communicate through the open literature or through more clandestine means, the essence of non-linear dynamics remains and it is very likely that chaotic behavior is the expected outcome. Then the question becomes how should we approach the regulation of inherently complex environments?